\newif\ifshow\showfalse

\def\Hc{H_{\rm C}} 
\def\Hs{H_{\rm S}} 

\def\1{\openone}

\def\seq#1{{\tt #1}}

\def\urlprefix#1#2{\hskip0pt plus0.01fil\discretionary{}{}{}%
  {\url{#2}}}

\def\urlprefix#1#2.{}

\documentclass[aps,pra,tightenlines,twocolumn,showpacs,floatfix]{revtex4}
\usepackage{amssymb}
\usepackage{epsf}
\usepackage[hypertex]{hyperref}
\begin{document}
\title{Soft-pulse dynamical decoupling in a cavity}
\author{Leonid P. Pryadko}
\affiliation{Department of Physics \& Astronomy, University of California,
  Riverside, California 92521, USA}
\author{Gregory Quiroz}
\affiliation{Department of Physics \& Astronomy, University of California,
  Riverside, California 92521, USA}
\date\today
\begin{abstract}
  Dynamical decoupling is a coherent control technique where the intrinsic and
  extrinsic couplings of a quantum system are effectively averaged out by
  application of specially designed driving fields (refocusing pulse
  sequences).  This entails pumping energy into the system, which can be
  especially dangerous when it has sharp spectral features like a cavity mode
  close to resonance.  In this work we show that such an effect can be avoided
  with properly constructed refocusing sequences. To this end we construct the
  average Hamiltonian expansion for the system evolution operator associated
  with a single ``soft'' $\pi$-pulse.  To second order in the pulse duration,
  we characterize a symmetric pulse shape by three parameters, two of which
  can be turned to zero by shaping.  We express the effective Hamiltonians for
  several pulse sequences in terms of these parameters, and use the results to
  analyze the structure of error operators for controlled Jaynes-Cummings
  Hamiltonian.  When errors are cancelled to second order, numerical
  simulations show excellent qubit fidelity with strongly-suppressed
  oscillator heating.
\end{abstract}
\maketitle
\section{Introduction} Quantum coherent control has found way into many
applications, including nuclear magnetic resonance (NMR), quantum information
processing (QIP), spintronics, atomic physics, etc.  The simplest control
technique is dynamical decoupling (DD), also known as refocusing.  The goal of
preserving coherence by averaging out the unwanted couplings is achieved most
readily by running precisely designed sequences of uniformly-shaped pulses
\cite{slichter-book,freeman-shaped-pulse,vandersypen-2004}.

In a closed system, the corresponding performance can be analyzed in terms of
the average Hamiltonian
theory\cite{waugh-huber-haeberlen-1968,waugh-wang-huber-vold-1968}.  To
leading order, the evolution over the refocusing period $\tau$ is indeed
described by the time-averaged Hamiltonian of the system in the ``rotating
frame'' defined by the control fields.  Generally, the average Hamiltonian is
constructed as a series in powers of $\tau$.  The number of the leading terms
of this expansion that are exactly zero determines the order $K$ of the
refocusing sequence.  Larger $K$ imply asymptotically more accurate
refocusing, with error terms scaling to zero faster with decreasing $\tau$.

For an open system, the dynamics associated with the bath degrees of freedom
can be also averaged out, as long as they are sufficiently slow.  With
leading-order ($K=1$) refocusing, the state decay processes are dramatically
suppressed\cite{kofman-kurizki-2001,kofman-kurizki-2004}, with a moderate
decrease of the dephasing rate\cite{pryadko-sengupta-kinetics-2006}, while
with second-order refocusing ($K=2$) both decay and dephasing are strongly
suppressed\cite{pryadko-sengupta-kinetics-2006}.

The decoherence analysis in Ref.\ \cite{pryadko-sengupta-kinetics-2006} was
based on the assumption of the low-frequency oscillator bath being near
thermal equilibrium.  This assumption becomes questionable if the bath has
sharp spectral features---e.g., if the controlled qubit system is coupled to a
local high-$Q$ oscillator.  On the other hand, such a situation where the
controlled system is coupled to an oscillator mode is quite common.  This
situation is realized in atomic physics, where the oscillator in question is
the cavity mode, while the continuous-wave (CW) excitation is used to suppress
the coupling\cite{villas-boas:041302}.  In several quantum computer designs,
nearly-linear oscillator modes are inherently present (e.g., mutual
displacement in ion
traps\cite{Cirac-Zoller-1995,vitali-tombesi-1999,you-2001,%
  kielpinski_2002:ion_trap,vitali-tombesi-2002}, or QCs based on electrons on
helium\cite{platzman-dykman-1999,%
  dykman-platzman-2000,lea-frayne-mukharsky,%
  dykman-platzman-seddighrad-2003}).  Finally, there are suggestions to
include local high-$Q$ oscillators in the QC designs to serve as ``quantum
memory'' \cite{pritchett:010301}, ``quantum information
bus''\cite{kapale:052304,wei:134506,geller:032311}, or as a part of the
measuring/control circuitry\cite{blais:032329}.

In this work we consider dynamical decoupling in a system where the spectral
function of the oscillator bath has a sharp resonance.  We include the
resonant mode and the corresponding couplings in the system Hamiltonian, and
consider the dynamics of the closed system driven by the refocusing pulses
applied to the qubits only.  We construct the average Hamiltonian for a
situation where one of the qubits is driven by a single symmetrical
one-dimensional $\pi$-pulse.  To second order, the expansion is characterized
by three parameters, two of which can be turned to zero by pulse shaping.  An
analysis of any refocusing sequence is then reduced to computing an ordered
product of evolution operators for individual pulses.  We illustrate the
technique by analyzing the controlled dynamics of a single qubit coupled to an
oscillator.  One of the analyzed sequences provides an order $K=2$ qubit
refocusing for any form of qubit--oscillator coupling.  The simulations done
for the Jaynes-Cummings Hamiltonian show excellent 
qubit fidelity with strongly-suppressed oscillator heating, as long as the
oscillator frequency bias exceeds the small coupling between the qubit and the
oscillator remaining in the effective Hamiltonian.  We argue that results of
Ref.~\onlinecite{pryadko-sengupta-kinetics-2006} for corresponding open system
remain applicable as long as this \emph{renormalized} coupling is small
compared to the resonance width.

\section{Background}

\subsection{Dynamical decoupling and effective Hamiltonian theory}

The main idea of dynamical decoupling is to drive the system in such a way as
to average out the effect of unwanted Hamiltonian couplings.  Obviously, this
only works if the control fields are large compared with the other terms of
the system Hamiltonian $\Hs$.

The easiest situation to analyze is where ``hard'' $\delta$-function
pulses are used.  In this case the system Hamiltonian can be ignored
altogether during the action of the pulse.  For a single
qubit, a $\pi$-pulse along the $x$-axis corresponds to the evolution
operator $X\equiv \exp(-i\pi s_x)=-i\sigma_x$, where $s_x=\sigma_x/2$
is the spin-$1/2$ operator.  The evolution operator for a sequence of
such pulses interrupting periods of free evolution can be written as a
product of the corresponding unitaries.  For example, the standard spin
echo\cite{hahn-1950} 
sequence of a $\pi$-pulse and a negative $\pi$-pulse in the $x$
direction followed by intervals of free evolution of equal
duration $\tau$ corresponds to the operator 
\begin{eqnarray}
  \label{eq:seq1}
  U_{X\text{--}\tau\text{--}\overline X\text{--}\tau\text{--}}&=&e^{-i
    \Hs\tau}\overline X \,e^{-i \Hs\tau} X .
\end{eqnarray}
Such expressions are easily simplified using the corresponding matrix algebra.
For the case of NMR, the system Hamiltonian is that of the chemical shift,
\begin{equation}
  \Hs={1\over2}\Delta \sigma_z,\label{eq:chem-shift}
\end{equation}
it anticommutes with the pulse unitary $X$, thus $
e^{-i \Hs\tau}\overline X=\overline Xe^{+i\Hs\tau}$, and the two-pulse
sequence (\ref{eq:seq1}) simplifies to the identity operator,
\begin{equation}
  \label{eq:seq1-ans}
  U_{X\text{--}\tau\text{--}\overline X\text{--}\tau\text{--}}=  
  \overline X e^{+i \Hs\tau} \,  e^{-i \Hs\tau} X=  \overline X  X=\1. 
\end{equation}
The simplicity of this formalism led to a number of strong
mathematical results applicable to refocusing with ideal
$\delta$-pulses.  In particular, a succession of ``concatenated''
refocusing sequences provide an excellent refocusing accuracy which
grows very rapidly with the number of pulses in a
sequence\cite{khodjasteh-Lidar-2005,khodjasteh-lidar-2007}.

In  practice, however, the hard-pulse condition may be 
difficult to satisfy, and one has to account for the corrections
associated with the action of the system Hamiltonian $\Hs$ during the
pulse.  If we denote the control Hamiltonian $\Hc(t)$, the total
Hamiltonian is
\begin{equation}
  \label{eq:ham}
  H(t)=\Hc(t)+\Hs.  
\end{equation}
The simplest 1st-order
decomposition (e.g., see 
Ref.~\cite{vandersypen-2004}) amounts to
adjusting the intervals of free evolution before and after the pulse, 
\begin{equation}
  \label{eq:decomp}
  e^{-i (\Hs+\Hc)\tau}\approx e^{-i  \Hs \tau_1} e^{-i \Hc\tau} e^{-i
    \Hs \tau_2}, 
\end{equation}
where $\Hc$ is assumed time-independent, 
with the precise value of $\tau_1$ and $\tau_2$ computed in order to
optimize the accuracy according to some fidelity measure.
Superficially, any combination such that $\tau_1+\tau_2=\tau$ appears
to provide equal accuracy to first order in $\tau$.  In fact, the
accuracy of the expansion relies on both $\Hs\tau$ and $\Hc\tau$ being
small; the results change non-trivially for finite-angle rotations.

A more systematic way to analyse the effect of pulse shape is in terms of the
average Hamiltonian
theory\hbox{\cite{waugh-huber-haeberlen-1968,waugh-wang-huber-vold-1968}}.
This is 
equivalent to constructing the cumulant expansion of the evolution operator
in powers of the system Hamiltonian in the interaction representation with
respect to the control Hamiltonian $\Hc(t)$ which is treated exactly.  

For a
system of qubits, the single-qubit control can be written most generally as
\begin{equation}
  \label{eq:ham-control}
   \Hc(t)= {1\over2}\sum_{n,\mu=(x,y,z)}  V^\mu_n(t)\,
  \sigma^\mu_n, 
\end{equation}
where $\sigma_n^\mu$, $\mu=x,y,z$, are the usual Pauli matrices for
the $n$-th qubit (spin).  We assume that the qubit
levels are nearly degenerate, or that Eq.~(\ref{eq:ham-control}) is
written in the rotating wave approximation with respect to the qubit
working frequency, so that the system Hamiltonian does not contain
large terms.
The control Hamiltonian~(\ref{eq:ham-control}) is a  sum of 
single-qubit terms, and the zeroth order evolution operator which obeys
the equation 
\begin{equation}
  \label{eq:zeroth-order}
  \dot U_0(t)=-i \Hc(t) U_0(t), \quad U_0(0)=\1, 
\end{equation}
can be constructed without much difficulty as a product of
corresponding single-qubit operators.  Then, the standard prescription
is to separate the fast dynamics due to control fields out of the
evolution operator by the decomposition, $U(t)=U_0(t)R(t)$, and write
the equation of slow evolution for $R(t)$,
\begin{equation}
  \dot R(t)=-i \tilde \Hs(t) R(t), \quad 
  \tilde \Hs (t)\equiv U_0^\dagger (t)\, \Hs \, U_0(t).
  \label{eq:evol-oper-inter}
\end{equation}
The system Hamiltonian in the interaction representation, $\tilde
\Hs (t)$, is small at the scale of the refocusing  period $\tau$, which
makes the time-dependent perturbation theory (TDPT) expansion
applicable.  The Magnus expansion, the exponentiated version of TDPT,
has somewhat better convergence properties.  It is written in terms of
cumulants $C_k$,
\begin{eqnarray}
  \label{eq:cumulant}
  R(t)\!&=&\!\exp\biglb( C_1(t)+C_2(t)+\cdots\bigrb), \\
  C_1(t)\!&=&\!-i\int_0^t dt_1 \tilde \Hs(t_1),\\ 
  C_2(t)\!&=&\!-{1\over 2}\int_0^t dt_2\int_0^{t_2} dt_1 [\tilde \Hs(t_1),\tilde \Hs (t_2)],\;\,\cdots .\,\,\,\,
\end{eqnarray}
Generally, the $k$-th cumulant $C_k$ contains a $k$-fold integration of
the commutators of the rotating-frame Hamiltonian $\tilde \Hs(t_i)$ at
different time moments $t_i$ and has an order $(t \Hs)^k$.  

Let us consider periodic dynamical decoupling with the period $\tau$,
$V_n^\mu(t+\tau)=V_n^\mu(t)$.  In addition, we request that the zeroth order
evolution operator~(\ref{eq:zeroth-order}) should also be periodic,
$U_0(\tau)=\1$ (zeroth-order refocusing condition).  Then, the Hamiltonian
$\tilde \Hs (t)$ is periodic, and the evolution operator over a time interval
commensurate with $\tau$, $\tau_n=n\tau$, can be factorized,
\begin{equation}
  \label{eq:factorized}
  R(n\tau)=[R(\tau)]^n=\exp(-in\tau H_{\rm ave}),
\end{equation}
where the ``average'' Hamiltonian $H_{\rm ave}$ is defined in terms of
cumulants, 
\begin{equation}
  \label{eq:average-ham}
  -i\tau H_{\rm ave}=C_1(\tau)+C_2(\tau)+\cdots.
\end{equation}
The main advantage of the average Hamiltonian theory is the improved
convergence at large time: regardless of the value of $t$, the TDPT
series needs to be convergent only at $|t|\le\tau$.

\subsection{Order of a refocusing sequence}

For order-$K$ refocusing, $K\ge1$, we require additionally that the first $K$
terms in the expansion of the effective Hamiltonian vanish, $C_1(\tau)= \ldots
= C_K(\tau)=0$.  For a closed system, a refocusing sequence of higher order
generally offers better scaling of accuracy with the sequence period.  If we
denote the maximum coupling in the system Hamiltonian as $J$ (more precisely,
$J=\| \Hs\|$, a norm of the system Hamiltonian), the $k$-th cumulant scales as
$C_k\sim(J\tau)^k$, while the corresponding term in the average Hamiltonian
$H_{\rm ave}^{(k)}\sim J^k \tau^{k-1}$.  Consequently, for order-$K$ dynamical
decoupling, the refocusing error defined as the norm of the deviation of the
unitary operator, $\delta_U=\|U(t)-\1\|$, scales as $\delta_U\sim (t/\tau)
(J\tau)^{K+1}$, while the corresponding fidelity scales as
$1-F\sim\delta_U^2\propto \tau^{2K}$.  Thus, in the same system, a refocusing
sequence of higher order can be run at a slower repetition rate.

Additional advantage of the effective Hamiltonian theory comes from its
locality.  The first-order effective Hamiltonian (cumulant $C_1$) contains only
the qubit interactions present in the original Hamiltonian $\Hs$.  The terms
$C_2$ contains connected pairs of such coupling terms.  Generally, if we
represent system Hamiltonian $\Hs$ as a graph, with qubits as vertices and
two-qubit couplings as corresponding edges, the expansion of the $k$-th order
cumulant $C_k$ can be represented as connected subgraphs with up to $k$ edges.
As a result, 1st-order refocusing condition, $C_1(\tau)=0$, can be verified by
analyzing clusters with up to two qubits; second-order refocusing, with the
additional condition $C_2(\tau)=0$, requires analysis of all clusters with up
to three qubits, etc.

This property, otherwise known as the cluster theorem\cite{Domb-Green-book},
allows one to design \emph{scalable} refocusing sequences whose refocusing
order is independent of the system size.  For
a linear chain with nearest-neighbor interaction, one can achieve this by
intermittently pulsing odd and even-numbered qubits.  Several
such particular sequences of order $K=2$ and higher were
demonstrated in Ref.~\onlinecite{sengupta-pryadko-ref-2005}.

\subsection{Sequence order and open-system refocusing}

Dynamical decoupling can be also effective against decoherence due to
low-frequency environmental modes\cite{viola-lloyd-1998}.  This can be
understood by noticing that the driven evolution with period
$\tau=2\pi/\Omega$ shifts some of the system's spectral weight by the
Floquet harmonics, $\omega\to \omega+n\Omega$.  With the first-order
average Hamiltonian for the closed system vanishing ($K=1$
refocusing), the original spectral weight at $n=0$ disappears
altogether, and the direct transitions with the bath degrees of
freedom are also suppressed as long as $\Omega$ exceeds the bath
cut-off frequency, $\Omega\agt\omega_c$
\cite{kofman-kurizki-2001,kofman-kurizki-2004}.  This corresponds to
effective suppression of dissipative ($T_1$) processes. 

The full analysis of decoherence, including both dissipative and reactive
($T_2$) processes, in the presence of order-$K$ dynamical decoupling, $K\le2$,
was done by one of the authors using the non-Markovian master equation in the
rotating frame defined by the refocusing fields
\cite{pryadko-sengupta-kinetics-2006}.  This involved a resummation of the
series for the Laplace-transformed resolvent of the master equation near each
Floquet harmonic, with subsequent summation of all harmonics.

The results of Ref.\ \cite{pryadko-sengupta-kinetics-2006} can be
summarized as follows.  With $K\ge1$ refocusing, there are no direct
transitions, which allows an additional expansion in powers of the
small adiabaticity parameter, $\omega_c/\Omega$.  In this situation
the decoherence is dominated by reactive processes (dephasing, or
phase diffusion).  With $K=1$, the bath correlators are modulated at
frequency $\Omega$.  This reduces the effective bath correlation time,
and the phase diffusion rate is suppressed by a factor
$\propto\omega_c/\Omega$.  With $K=2$ refocusing, all 2nd-order terms
involving correlators of the bath coupling at zero frequency,
$\omega=0$, are cancelled.  Generically, this leads to a suppression
of the dephasing rate by an additional factor
$\propto(\omega_c/\Omega)^2$, while in some cases (including
single-qubit refocusing) all terms of the expansion in powers of the
small adiabaticity parameter $(\omega_c/\Omega)$ disappear.  This
causes an \emph{exponential} suppression of the dephasing rate, so
that an excellent refocusing accuracy can be achieved with relatively
slow refocusing, $\Omega\agt\omega_c$.

\section{Dynamical decoupling of a generically-coupled qubit} 
The analysis in Ref.~\onlinecite{pryadko-sengupta-kinetics-2006} was done for
a generic thermal bath with a featureless quasi-continuous spectrum
characterized by the upper cut-off frequency $\omega_c$.  The absence of sharp
features justified an approximation where bath memory effects were essentially
ignored at the scale of the decoherence time, although bath correlations at
shorter times are crucial for describing the effects of dynamical decoupling.
A sharp spectral feature, like a high-$Q$ cavity mode, makes the starting
point of the analysis\cite{pryadko-sengupta-kinetics-2006}, the non-Markovian
master equation involving only qubits, questionable, unless the decoherence
time {\em in the absence\/} of refocusing is long compared with the
equilibration time of the high-$Q$ mode.

\subsection{Model}
In this work we include any sharp quantum mode(s) into the ``system'' part
of the Hamiltonian $\Hs$, and consider the driven quantum dynamics of the
resulting closed system.  
Compared with a system of qubits, a quantum oscillator admits a wider
variety of linear or non-linear couplings.  By this reason we begin
with a model of a qubit with most general couplings,
\begin{equation}
  \Hs={\sigma_x}A_x+{\sigma_y}A_y+{\sigma_z}A_z+A_0, 
  \label{eq:single-qubit}  
\end{equation}
where $\sigma_\mu$ are the qubit Pauli matrices and $A_\nu$, $\nu=0,x,y,z$ are
the operators describing the degrees of freedom of the rest of the system
which commute with $\sigma_\mu$, $[\sigma_\mu,A_\nu]=0$ but not necessarily
with each other.
\subsection{Pulse structure to linear order}
First, consider the qubit evolution driven by a one-dimensional pulse, 
\begin{equation}
  \Hc=\textstyle{1\over2}{\sigma_x}V_x(t), \quad 0<t<\tau_p,
  \label{eq:one-pulse}  
\end{equation}
where the field $V_x(t)$ defines the pulse shape.  The unitary evolution
operator to zeroth order in $\Hs$ is 
simply
\begin{equation}
  U_0(t)=e^{-i\sigma_x \phi(t)/2}, \quad \phi(t)\equiv \int_0^t
  dt'\,V_x(t').\label{eq:control-evolution}   
\end{equation}
When acting on the spin operators, this is just a rotation, e.g.,
$U_0(t)\sigma_yU_0^\dagger (t)=\sigma_y\cos \phi(t)+\sigma_z \sin \phi(t)$.

For inversion pulses with the net rotation angle
$\phi(\tau_p)=\pi$, with a symmetric  shape,
$V_x(\tau_p-t)=V_x(t)$, the average of the cosine over pulse duration is zero
by symmetry,
\begin{equation}
  \langle
  \cos\phi(t)\rangle_p\equiv {1\over \tau_p}\int_0^{\tau_p}dt \cos\phi(t)=0.
\label{eq:cos-average}
\end{equation}
In such a case, the first two terms of the expansion, $X=X^{(0)}+\tau_p
X^{(1)}+\tau_p^2X^{(2)}+\ldots$, of the unitary evolution operator $X\equiv
U(\tau_p)$ in powers of pulse duration $\tau_p$ read: 
\begin{eqnarray}
  \label{eq:pi-expansion-0th}
  X^{(0)}&=&-i\sigma_x,\\
  \label{eq:pi-expansion-1st}
  X^{(1)}&=& -
   A_x-\sigma _x A_0+is
    \left( \sigma_y A_y+ \sigma_z A_z\right).
\end{eqnarray}
Here the dimensionless parameter 
\begin{equation}
  s\equiv \langle \sin \phi(t)\rangle_p\label{eq:s-defined}  
\end{equation}
is the only one that characterizes the pulse shape in this order. 

We note that for $\pi$-pulses the sine of evolution angle is non-zero
only over the duration of the pulse; the time intervals before the
beginning and after the end of the pulse where $V_x(t)=0$ do not
contribute to the value of~$s$.  Because of that, this parameter can
be viewed as a measure of the effective duration of the pulse.  For
example, while $s=0$ for an infinitely short $\delta$-pulse, for a
Gaussian pulse\cite{bauer-gauss} of width $\tau$,
\begin{equation}
  G_\tau(t+\tau_p/2)\equiv {\pi^{1/2}\over\tau}e^{-t^2/\tau^2}, \quad
  \tau\ll \tau_p, 
  \label{eq:gauss} 
\end{equation}
this parameter is $s\approx 1.5\tau/\tau_p$ (see Tab.~\ref{tab:params}).  

To create a pulse shape with effectively zero width to linear order, one can
compensate for positive values of $\sin\phi(t)$ near the middle of the
interval by making $V_x(t)$ somewhat negative near the beginning and the end
of the interval.  Such a first-order self-refocusing pulse was first suggested
by Warren \cite{warren-herm} as a ``Hermitian'' shape,
\begin{equation}
  \label{eq:herm}
  H_\tau(t+\tau_p/2)\equiv G_\tau(t+\tau_p/2){1-\gamma\,
 t^2/\tau^2\over
    1-\gamma/2}, 
\end{equation}
where the precisely computed value $\gamma=0.9609317217$ is somewhat different
from that in Ref.~\onlinecite{warren-herm}.  The corresponding fixed-length
1st-order self-refocusing pulse shapes $S_L$, with all derivatives up to and
including $(2L-1)\,$st vanishing at the ends of the interval, $L=1,2$, were
constructed\cite{sengupta-pryadko-ref-2005} 
in terms of their Fourier
coefficients\cite{geen-freeman-burp},
\begin{equation}
  V(t+\tau_p/2) =A_0+\sum_m A_m \cos (m\Omega_p t),
  \label{eq:pulse-fourrier}
\end{equation}
where the angular frequency $\Omega_p=2\pi/\tau_p$ is related to the
full pulse duration $\tau_p$; the coefficients $A_m$ are listed in
Ref.~\onlinecite{sengupta-pryadko-ref-2005}.
Compared to Hermitian pulse shape, the main
advantage of pulses $S_L$ is their smaller power [the maximum amplitude of the
field $V(t)$].

For 1st-order self-refocusing pulse shapes such as $H_\tau$ or $S_L$, the
terms proportional to $s$ in Eq.~(\ref{eq:pi-expansion-1st})
disappear, and the expansion of the unitary
operator simplifies~to
\begin{eqnarray}
  \label{eq:pi-expansion-1st-b}
  X_{s=0}=-i \sigma _x-\tau_p \bigl( A_x+\sigma _x A_0\bigr)
  +\mathcal{O}(\tau_p^2). 
\end{eqnarray}

\subsection{Pulse structure to quadratic order}
The structure of the pulse to quadratic order is easily computed as the next
order in the TDPT.
We have
[cf.~Eqs.~(\ref{eq:pi-expansion-0th}), (\ref{eq:pi-expansion-1st})],
\begin{eqnarray}
    X^{(2)}
    &=&  \nonumber 
  + {i\over 2}
  \biglb(
  \{A_0 , A_x\}    + 
  \sigma_x  (A_0^2+A_x^2)\bigrb) 
       \\ & & \hskip-0.2in\nonumber  
  + 
  \zeta \bigl( [A_0 , \sigma_y A_z-\sigma_z A_y] + i 
  \{A_x, \sigma_y A_y+\sigma_z A_z\}\bigr)
  \\ & & \hskip-0.2in\nonumber
  +
  {s\over 2}\bigl( \{A_0,\sigma_y A_y+\sigma_z A_z\}-i[A_x,\sigma_y
  A_z-\sigma_z A_y]\bigr)
  \\ & &      \hskip-0.2in\nonumber 
  + 
  \alpha  \bigl(A_y^2+A_z^2+ i  \sigma_x  [A_y ,  A_z]
  \bigr), 
  \\ & & \hskip-0.2in
  +
  {s^2\over 2}\biglb([A_z,A_y]+{i\sigma_x}(A_y^2+A_z^2)\bigrb),
  \label{eq:pi-expansion-2nd}
\end{eqnarray}
where we parametrized the pulse shape in terms of two additional parameters 
\begin{eqnarray}
\label{eq:alpha-defined}
  \alpha&\equiv &\langle
\theta(t-t')\sin [\phi(t)-\phi(t')]\rangle_p, \\
\label{eq:zeta-defined}\zeta&\equiv&
\langle\theta(t-t') \cos \phi(t')\rangle_p ,
\end{eqnarray}
with the two-time averages
over pulse duration 
\begin{equation}
  \langle
  f(t,t')\rangle_p\equiv \int_0^{\tau_p} {dt\over \tau_p}
\int_0^{\tau_p} {dt'\over \tau_p} f(t,t').
\label{eq:pulse-average}
\end{equation}
We also use the notations $[A,B]\equiv AB-BA$ for the commutator and
$\{A,B\}=AB+AB$ for the anticommutator of two operators.

The effect of the parameter $\alpha$ was studied previously in
Ref.~\onlinecite{sengupta-pryadko-ref-2005}.  The condition $\alpha=0$ is
necessary to obtain NMR-style one-dimensional second-order self-refocusing
pulses.  
Indeed, the corresponding system Hamiltonian is just that of the
chemical shift, Eq.~(\ref{eq:chem-shift}), 
thus $A_x=A_y=A_0=0$ and
$A_z=\Delta/2$ in Eq.~(\ref{eq:single-qubit}).  Then, the evolution operator
to quadratic order in $\bar\tau\equiv \tau_p \Delta/2$ is simply
\begin{equation}  
X_\mathrm{\Delta}=-i\sigma_x+is\bar\tau \sigma_z +\bar\tau^2\bigl(\alpha+i
{\sigma_xs^2/ 2}\bigr)+\mathcal{O}(\bar\tau^3); \label{eq:chemical-shift}
\end{equation}
the linear and quadratic corrections disappear entirely for second-order
self-refocusing pulses such that $\alpha=s=0$.   
These are exactly the conditions used to design pulse shapes $Q_L$
(Ref.~\onlinecite{sengupta-pryadko-ref-2005}). The index $L=1,2$ denotes
the parameter for the additional
condition that the first $2L$ derivatives vanish at the ends of the
interval, $V^{(l)}(0)=V^{(l)}(\tau_p)=0$, $l=0,\ldots 2L-1$.

The actual values of the parameters $s$, $\alpha$, and $\zeta$ for several
pulse shapes are listed in Tab.~\ref{tab:params}.  We note that for an ideal
$\delta$-pulse, the parameters $s=\alpha=0$, whereas $\zeta=1/4$ is not
particularly small.  For all ``soft'' pulse shapes listed, the values of
$\zeta$ are quite close to this value.  The values of the parameter $\alpha$
are numerically small for all 1st-order self-refocusing pulses with $s=0$.
The second-order self-refocusing pulses $Q_1$, $Q_2$ with $s=\alpha=0$ may
work as a perfect replacement of $\delta$-pulses to second order
accuracy\cite{sengupta-pryadko-ref-2005}.

\begin{table}[hc]
  \centering
  \begin{tabular}[c]{c|c|c|c|}
    pulse & $s$ & $\alpha/2$ & $\zeta$ \\ 
    \hline 
    $\pi\delta(t-\tau_p/2)$ & 0 & 0 & 1/4 \\ 
    $G_{0.05}$ & 0.0744895 & 0.0349708 & 0.249476 \\ 
    $G_{0.10}$ & 0.148979  & 0.0653938 & 0.247905 \\ 
    $H_{0.05}$ & 0  & 0.00153849& 0.249647\\
    $H_{0.10}$ & 0  & 0.00615393&  0.248589\\
    $S_1$ & 0 & 0.0332661 & 0.238227 \\
    $S_2$ & 0 & 0.0250328 & 0.241377 \\ 
    $Q_1$ & 0 & 0 & 0.239889 \\     
    $Q_2$ & 0 & 0 & 0.242205 
  \end{tabular}
  \caption{Parameters of several symmetric pulse shapes.  The first line 
    represents  the ``hard'' $\delta$-function pulse, $G_{0.05}$ denotes the
    Gaussian\cite{bauer-gauss} pulse with the width $\tau=0.05\tau_p$, see
    Eq.~(\protect\ref{eq:gauss}), $H_{0.05}$ is the 
    corresponding Hermitian\cite{warren-herm} pulse, see
    Eq.~(\protect\ref{eq:herm}), while $S_L$ and $Q_L$ denote 
    the 1st and 2nd-order self-refocusing pulses from
    Ref.~\cite{sengupta-pryadko-ref-2005}, with up to $2L-1\,$st derivative
    vanishing at the ends of the interval of duration $\tau_p$.}
  \label{tab:params}
\end{table}

\section{Common pulse sequences.}
  Transforming
Eqs.~(\ref{eq:pi-expansion-1st}), (\ref{eq:pi-expansion-2nd})
appropriately, we can now easily compute the result of application of
any pulse sequence.  In particular, the $\pi$-pulse $\overline X$
applied along the $-x$ direction can be obtained from $(-X)$ with the
substitution $\alpha\to-\alpha$.  As a result, e.g., the expansion of
the evolution opeator for the one-dimensional sequence $\overline X X$
[cf.\ Eq.~(\ref{eq:seq1})] can be written as
\begin{eqnarray}
  \nonumber 
  \lefteqn{{\overline X X}= \openone -2 i \tau_p \biglb( A_0 + \sigma_x
    A_x-s(\sigma_y 
    A_z-\sigma_z A_y)\bigrb)}& & \hskip0.9\columnwidth\\ & &
  -2 \tau_p^2 \biglb(A_0 +\sigma_x A_x-s(\sigma_y
  A_z-\sigma_z A_y)\bigrb)^2 +\mathcal{O}(\tau_p^3),\label{eq:x-bar-x}
\end{eqnarray}
or it can be re-exponentiated as evolution over time interval
$2\tau_p$ with the average Hamiltonian
\begin{equation} 
  H_{ \overline X X}=
  A_0 +\sigma_x  A_x-s(\sigma_y
  A_z-\sigma_z A_y)+\mathcal{O}(\tau_p^2). \label{eq:1d-2pulse}
\end{equation}
We can attempt to correct for the terms proportional to $s$ by using a longer
sequence, e.g., $ X \overline X \overline X X$.  However,
while the term is corrected in the leading-order effective Hamiltonian, we
acquire a correction in the next order,
\begin{eqnarray} 
  \nonumber 
  H_{X \overline X \overline X X}&=&
  A_0 +\sigma_x  A_x- s\tau_p\bigl\{A_x,\sigma_y
  A_y+\sigma_zA_z\bigr\}\\ & & 
  +is \tau_p [A_0,\sigma_yA_z-\sigma_z A_y]+\mathcal{O}(\tau_p^2).
  \label{eq:1d-4pulse} 
\end{eqnarray}

While the external Hamiltonian $A_0$ cannot be averaged out by acting on the
qubit, the term proportional to $\sigma_x$ can be also suppressed with the
help of two-dimensional sequences.  The expression for the unitary
$U_Y(\tau_p)\equiv Y$ resulting from application of the pulse along the
$y$-direction can be easily obtained from Eqs.~(\ref{eq:pi-expansion-1st}),
(\ref{eq:pi-expansion-2nd}) by cyclic permutation of indices. Then, for
example, the refocusing sequence $\mathbf{4p}\equiv \mathbf{4p}({xy})\equiv
X\overline Y XY$ corresponds to the effective Hamiltonian
\begin{eqnarray}
  \nonumber  
   H_{{\bf 4p}}&=&A_0+{s\over2}(\sigma_x A_z-\sigma_z
   A_y)-{i\tau_p\over2} [A_0,\sigma_x A_x-\sigma_y A_y] 
   \\   & &\hskip-0.3in 
   -\tau_p{\alpha\over
    2}\sigma_y(A_x^2+A_z^2) + \tau_p{i\alpha\over2}[A_z,A_y]
  \nonumber   \\ & &\hskip-0.3in    
  -\tau_p{1+4\zeta\over4}\sigma_z \{A_x,A_y\}
  +\mathcal{O}(\tau_p^2,s\tau_p),
  \label{eq:4p}
\end{eqnarray}
where we dropped linear in $\tau_p$ terms proportional to $s$.

The symmetric 8-pulse sequence ${\bf 8s}={YX\overline Y XX\overline Y XY}$
produces the effective Hamiltonian
\begin{eqnarray}
  H_{\mathbf{8s}}\!&=&\!A_0
  +s\tau_p\Biglb({i\over4}[A_z,A_x+A_y]+{1\over2}(\sigma_x A_y^2-\sigma_y
  A_x^2)\nonumber\\ & & \quad
  +{1\over4}\sigma_y\{A_x,A_y\}+{1\over4}\sigma_z\{A_y, A_z\}  \nonumber\\ & &
  \quad  
  +{i\over2}\bigl[A_0,\sigma_y A_z+\sigma_z A_x+{3\over2} \sigma_z
  A_y-{5\over2}\sigma_x
  A_z\bigr]
\Bigrb)\nonumber\\ & & 
-{\alpha\tau_p\over 2}
    \Bigl(\sigma_y (A_x^2+A_z^2)+i[A_y,A_z]\Bigr)
    +\mathcal{O}(\tau_p^2),
    \qquad 
  \label{eq:8s}
\end{eqnarray}
while the antisymmetric sequence ${\bf 8a}\equiv {\overline Y\overline X
  Y\overline XX\overline Y XY}$
corresponds to 
\begin{equation} 
   H_{{\bf 8a}}=A_0+{s\over 2}(\sigma_x A_z-\sigma_z A_y)
    +\mathcal{O}(\tau_p^2). 
  \label{eq:8a}
\end{equation}
We note that in the latter case there is a leading-order term proportional to
$s$ but no terms in order $\tau_p$; this sequence produces 2nd-order
refocusing already with 1st-order pulses.

\section{Qubit in a cavity}
To illustrate these results,  consider a qubit placed in a lossless
cavity with a single mode nearly-resonant with the qubit.  We consider the
simplest case where the system can be described by the Jaynes-Cummings
Hamiltonian,
\begin{equation}
  \label{eq:jc-ham}
  \Hs=\omega_r b^\dagger b+{\omega_0 \over 2}\sigma^z-g(b^\dagger
  \sigma_-+\sigma_+b),  
\end{equation}
where $\sigma_{\pm}\equiv (\sigma_x\pm i \sigma_y)$, and $\omega_r$ and
$\omega_0$ are the frequency biases for the cavity and the qubit respectively.
Eq.~(\ref{eq:jc-ham}) can be also written in the form~(\ref{eq:single-qubit})
with $A_x=-g (b+b^\dagger)/2$, $A_y=ig (b^\dagger-b)/2$, $A_z=\omega_0$, and
$A_0=\omega_r\, b^\dagger b$.  We concentrate on the special case $\omega_0=0$
which corresponds to working in the ``rotating frame,'' with the control
fields [Eq.~(\ref{eq:ham-control})] applied on resonance with the qubit.  Note
that we assume the control fields to be applied directly at the qubit and not
at the oscillator\cite{blais:032329}; for a linear oscillator the
corresponding compensation can be achieved by a spectral filter.
\subsection{Sequence 4p}
\begin{figure}[tp]
  \centering
  \epsfxsize=1.00\columnwidth
  \epsfbox{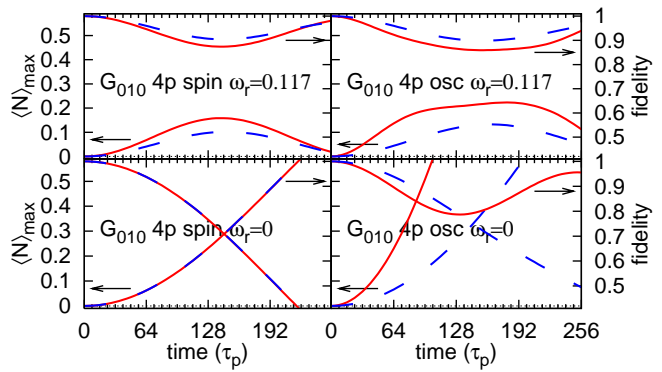}
  \caption{(color online) Evolution of the qubit fidelity and the number of
    quanta at the oscillator $\langle n\rangle$ (minimized and maximized over
    initial qubit states, respectively) at the end of the refocusing interval.
    Length-4 sequence \seq{4p} with Gaussian pulses $G_{010}$ applied on
    resonance with the qubit, $\omega_0=0$ in Eq.~(\protect\ref{eq:jc-ham}).
    Left panels: 
    oscillator restricted to $n=0,1$ states, which makes it effectively a
    qubit.  Right panels: oscillator restricted to $n\le 8$ levels, which over
    the simulation time is equivalent to infinity.  Bottom panels correspond
    to oscillator in resonance with the qubit, top panels correspond to
    oscillator frequency bias $\omega_r=0.117 (2\pi/\tau_p)$.  Red solid
    lines correspond to control pulses applied along $X$ and $Y$ axes
    [sequence \textbf{4p}(xy)]; blue dashed lines correspond to control pulses
    applied along $X$ and $Z$ axes [sequence \textbf{4p}(xz)].}
  \label{fig:4p_G010}
\end{figure}

When the sequence \textbf{4p} 
is used with a Gaussian or other
pulse shape with $s\neq0$, the effective Hamiltonian to leading order can be
written as
\begin{equation}
  \label{eq:cavity-1}
  H_{\bf 4p}=\omega_r b^\dagger b + 
  {s\omega_0\over2}\sigma_x+ {isg\over4}\sigma_z ( b^\dagger -b )+\mathcal{O}(\tau_p).
\end{equation}
The original exchange-like oscillator coupling in Eq.~(\ref{eq:jc-ham}) is
replaced by the qubit phase coupling in Eq.~(\ref{eq:cavity-1}).  While the
coupling magnitude is reduced by a small factor $\propto s$, it does not go
down with more frequent pulse application.  The same holds true if the
oscillator is replaced by an auxiliary qubit (left panel in
Fig.~\ref{fig:4p_G010}), or if the pulses are applied in the $x$-$z$ direction
[dashed lines in Fig.~\ref{fig:4p_G010}; the effective Hamiltonian is given by
Eq.~(\ref{eq:cavity-2xz}) below].  One can see from Fig.~\ref{fig:4p_G010}
that the refocusing accuracy is more or less similar for all these cases.

\begin{figure}[htbp]
  \centering
  \epsfxsize=1.00\columnwidth
  \epsfbox{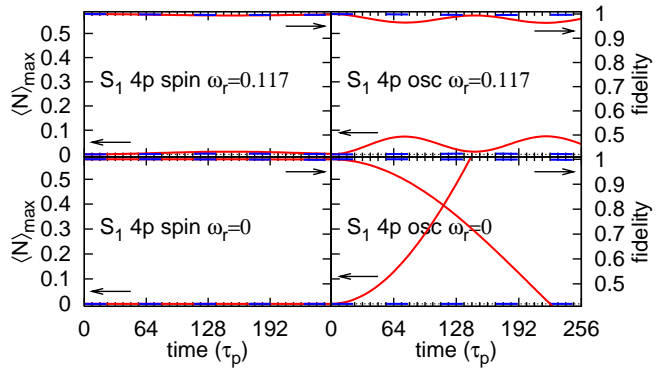}
  \caption{(color online) As in Fig.~\ref{fig:4p_G010} but with 1st-order
    self-refocusing pulses S$_1$.  The corresponding curves for pulses Q$_1$
    (not shown) are virtually identical with linear scale.}
  \label{fig:4p_S1}
\end{figure}

When the sequence \textbf{4p} is applied with 1st-order self-refocusing pulses
(Hermitian or $S_L$), the first-order terms in the effective Hamiltonian are
gone.  This leaves terms linear in $\tau_p$ as the leading-order correction
\begin{eqnarray}
  \nonumber 
  H_{\bf 4p}&=&\omega_r b^\dagger b -i{\tau_p g\omega_r\over
    4}\sigma_x(b^\dagger -b) \nonumber  \hspace{1.3in} \\ 
   & & 
   -i{1+4\zeta\over8}\tau_p g^2 \sigma_z (b^2-\mathrm{h.c.})
   +\mathcal{O}(s,\alpha\tau_p,\tau_p^2)
  \label{eq:cavity-2}
\end{eqnarray}
where we set $s=0$ and dropped the relatively small term $\propto \alpha
\tau_p$.  We note that the first two terms in Eq.~(\ref{eq:cavity-2})
disappear with $\omega_r=0$.  The third term is suppressed when the oscillator
is replaced by an auxiliary qubit ($b\to\tau_-$, $b^\dagger\to \tau_+$,
$b^\dagger b\to \tau_z/2$).  The same happens if the sequence is applied in
the $x$-$z$ direction [sequence $\mathbf{4p}(xz)=X\overline Z XZ$], with the
corresponding effective Hamiltonian
\begin{eqnarray}
  H_{\bf 4p}^{(xz)}& =&\omega_r b^\dagger b+{isg\over4} \sigma_x
  (b^\dagger-b) \nonumber \\ & & 
 + i{\tau_p g\omega_r \over 4}\sigma_x
  (b^\dagger-b) +
  \mathcal{O}(s\tau_p,\alpha\tau_p,\tau_p^2). 
  \label{eq:cavity-2xz}
\end{eqnarray}
The net effect is
that the refocusing accuracy is dramatically improved when the oscillator is
replaced by an auxiliary qubit, or when the four-pulse sequence is applied in
the $x$-$z$ plane, with the additional mode being in resonance with the qubit.
In such cases, the only remaining linear in $\tau_p$ terms in the effective
Hamiltonian are those proportionaly to the small parameter $\alpha$, and the
refocusing gets much more accurate, see Fig.~\ref{fig:4p_S1}.

\subsection{Sequence 8a}

\begin{figure}[tp]
  \centering
  \epsfxsize=1.00\columnwidth
  \epsfbox{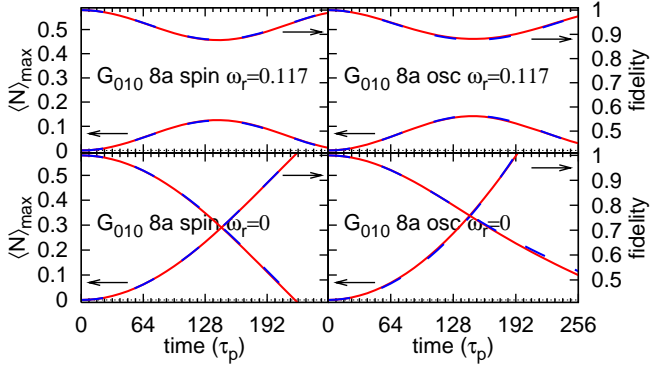}  
  \caption{(color online) As in Fig.~\ref{fig:4p_G010} but for the sequence
    {\bf 8a}.  The term $\propto s$ in Eq.~(\ref{eq:cavity-8a}) produces large
    errors comparable for those for \textbf{4p} sequence in
    Fig.~\ref{fig:4p_G010}.  This indicates that this sequence is less stable
    to pulse shape errors as, e.g., the symmetric sequence \textbf{8s}.}
  \label{fig:8a_G010}
\end{figure}

\begin{figure}[hbp]
  \centering
  \epsfxsize=1.00\columnwidth
  \epsfbox{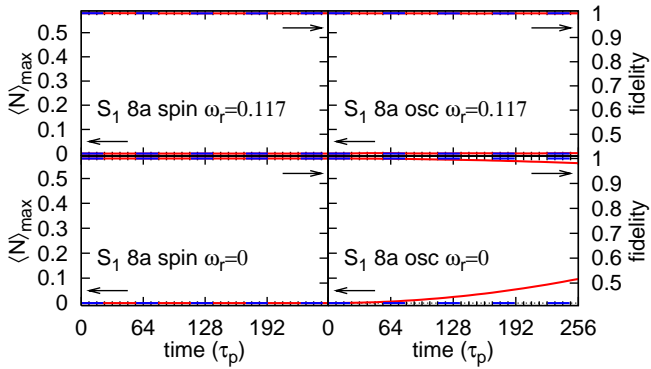}  
  \caption{(color online) As in Fig.~\ref{fig:4p_G010} but for the sequence
    {\bf 8a} with pulses S$_1$.  The corresponding curves for pulses Q$_1$
    (not shown) are virtually identical with linear scale.}
  \label{fig:8a_S1}
\end{figure}

The antisymmetric eight-pulse sequence \textbf{8a} produces the
effective Hamiltonian
\begin{equation}
  \label{eq:cavity-8a}
  H_\mathbf{8a}=\omega_r b^\dagger b+
   {isg\over4}\sigma_z ( b^\dagger -b )
  +\mathcal{O}(\tau_p^2).
\end{equation}
The refocusing is 1st-order with Gaussian pulses (Fig.~\ref{fig:8a_G010}),
with the errors comparable to those of the sequence \textbf{4p} (cf.\
Fig.~\ref{fig:4p_G010}).  With self-refocusing pulses S$_1$
[Fig.~\ref{fig:8a_S1}] or Q$_1$ (not shown) the sequence produces 2nd-order
refocusing.  With all 1st and 2nd-order error terms cancelled, the refocusing
accuracy is improved substantially.


\subsection{Sequence 8s}
\begin{figure}[tp]
  \centering
  \epsfxsize=1.00\columnwidth
  \epsfbox{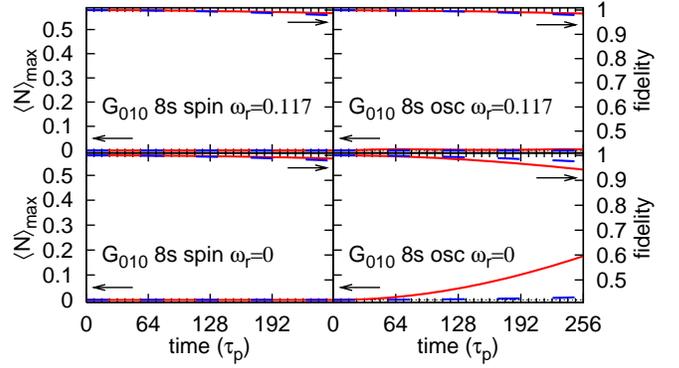}  
  \caption{(color online) As in Fig.~\ref{fig:4p_G010} but for the sequence {\bf 8s}.}
  \label{fig:8s_G010}
\end{figure}

\begin{figure}[tp]
  \centering
  \epsfxsize=1.00\columnwidth
  \epsfbox{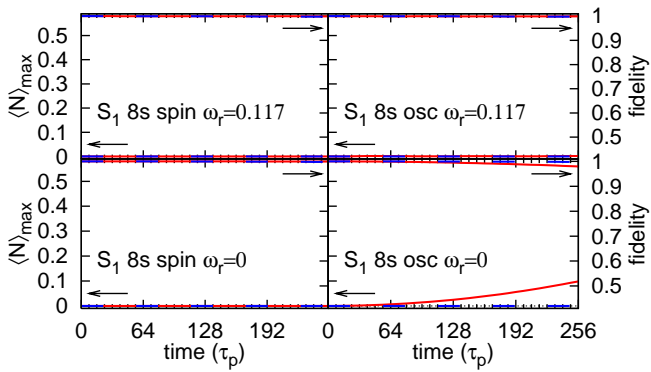}  
  \caption{(color online) As in Fig.~\ref{fig:4p_G010} but for the sequence
    {\bf 8s} with pulses S$_1$.  Formally, the sequence is of the same order
    as with Gaussian pulses, see Fig.~\ref{fig:8s_G010}; the noticeable
    differences are due to error terms of higher order which we dropped in our
    calculations.  The off-resonance refocusing is excellent (the fidelity and
    $\langle N\rangle$ curves in the top right panel run along the
    corresponding axes), while the on-resonance performance is also good. The
    plots with second-order pulses Q$_1$ (not shown) look almost identical.
    Note also that with the linear scale of these plots, the curves here are
    almost indistinguishable from those in 
    Fig.~\ref{fig:8a_S1}.}
  \label{fig:8s_S1}
\end{figure}

The symmetric eight-pulse sequence \textbf{8s} produces the effective
Hamiltonian 
\begin{equation}
  \label{eq:cavity-3}
  H_\mathbf{8s}=\omega_r b^\dagger b-{\alpha g^2 \tau_p\over 8}\sigma_y
  (b^\dagger +b)^2+\mathcal{O}(s\tau_p,\tau_p^2)
\end{equation}
which corresponds to Eq.~(\ref{eq:cavity-2}) with the larger terms already
suppressed.  The refocusing is first-order with either Gaussian
[Fig.~\ref{fig:8s_G010}] or 1st-order pulses [see Fig.~\ref{fig:8s_S1} with
pulses S$_1$], and second-order with second-order pulses $Q_L$ (not shown).
With the linear scale of our plots, there is only a slight difference between
zeroth and first order pulses (due to higher-order terms), and no visible
difference between pulses S$_1$ and Q$_1$.

We note that the performance of this sequence is very close to that of the
antisymmetric sequence \textbf{8a}.  Nevertheless, the symmetric sequence
\textbf{8s} provides better stability with respect to the pulse shape errors,
and it is also expected to result in better visibility (initial decoherence in
Ref.~\cite{pryadko-sengupta-kinetics-2006}) when used in an open system.

\subsection{Discussion}

Our simulations show that with properly designed refocusing, to an excellent
accuracy, the quantum oscillator coupled to the qubit remains in the ground
state, while the qubit fidelity remains very close to unity.  Numerical
results agree with the predictions from the analytically computed effective
Hamiltonians which show greately reduced coupling between the qubit and the
oscillator.  Then, in the closed system, a very small frequency bias (of order
of renormalized coupling value) becomes sufficient to effectively disconnect
the oscillator mode and protect the qubit coherence.  Similar effect can be
achieved when the finite width $\Gamma$ of the resonance is taken into
account: for refocusing to be effective, $\Gamma$ must be large compared with
the renormalized qubit coupling.

In other words, refocusing is effective as long as the renormalized qubit
coupling with the resonant mode is small on the scale of the resonance width.
This gives the modified, more broadely applicable, criterion for applicability
of the results of Ref.~\cite{pryadko-sengupta-kinetics-2006}.  In particular,
we expect the second-order sequence \textbf{8s} with 2nd-order pulses Q$_n$ to
provide an excellent refocusing accuracy to system Hamiltonian of
Jaynes-Cummings form~(\ref{eq:jc-ham}) also in the presence of a thermal bath,
as long as the refocusing rate is sufficiently high.  In addition to the
condition on renormalized value of the high-$Q$ oscillator mode coupling, the
refocusing period $\tau$ must be below the threshold value of order of the
inverse bath cut-off frequency refocusing frequency, $\omega_c^{-1}$, see
Ref.~\onlinecite{pryadko-sengupta-kinetics-2006}.

\section{Conclusions.} 

In this work we analyzed the performance of soft-pulse dynamical decoupling in
the presence of a sharp resonance mode.  Because of the associated memory
effects, such a problem cannot be addressed by considering master-equation
dynamics for the qubit-system density matrix alone.  Instead, we included the
resonant mode and the corresponding couplings in the system Hamiltonian, and
considered the dynamics of the resulting closed quantum system driven by a
sequence of soft or hard refocusing pulses applied to the qubit.  The analysis
was done in terms of the effective Hamiltonian theory which describes the
evolution of the system ``stroboscobically'' at the time moments commensurate
with the refocusing period.

In fact, to make our results applicable to a large number of possible coupling
terms between the qubit and the oscillator, we solved a more general problem
of an arbitrary coupled [see Eq.~(\ref{eq:single-qubit})] controlled qubit.
The main result of this work is the expansion of the unitary evolution
operator (\ref{eq:pi-expansion-1st}), (\ref{eq:pi-expansion-2nd}) and the
classification of the corresponding parameters in Tab.~\ref{tab:params}.  This
allows an explicit computation of the error operators associated with
refocusing in systems of arbitrary complexity.

We also computed the effective Hamiltonians for several single-qubit
refocusing sequences.  To quadratic order, all coupling terms are cancelled if
the length-8 sequences \textbf{8a} [Eq.~(\ref{eq:8a})] or \textbf{8s}
[Eq.~(\ref{eq:8s})] are used with 1st or 2nd-order self-refocusing pulses
respectively.  We illustrated the general analytical results on the specific
example of the driven Jaynes-Cummings Hamiltonian.  The results of simulations
agree with the predictions based on the analytically computed second-order
effective Hamiltonians, although in some cases the effects of higher-order
terms not included in the calculation are noticeable.

For robust single-qubit refocusing, we recommend the symmetric 8-pulse
sequence \textbf{8s} applied with second-order self-refocusing pulses.  This
sequence provides excellent refocusing accuracy for any form of the coupling
of the qubit with outside world, and it is stable with respect to pulse shape
errors.  As the second-order symmetric sequence, it should also work well for
open systems, as long as the environment is slow on the scale of the
refocusing rate\cite{pryadko-sengupta-kinetics-2006}.  The thermal bath is
expected to remain close to equilibrium, and refocusing to perform well, as
long as sharp features in the spectral function are wide on the scale of the
\emph{renormalized} value of the corresponding couplings.

\section{Acknowledgements.} This research was supported in part by the NSF
grant No.\ 0622242 (LP) and the Dean's Undergraduate Research fellowship at
UCR (GQ).


\end{document}